# Graphene-like Membranes: From Impermeable to Selective Sieves

G. Brunetto and D. S. Galvao
Applied Physics Department, State University of Campinas, 13083-970, Campinas-SP, Brazil.

## ABSTRACT

Recently, it was proposed that graphene membranes could act as impermeable atomic structures to standard gases. For some other applications, a higher level of porosity is needed, and the so-called Porous Graphene (PG) and Biphenylene Carbon (BPC) membranes are good candidates to effectively work as selective sieves. In this work we have used classical molecular dynamics simulations to study the dynamics of membrane permeation of He and Ar atoms and possible selectivity effects. For the graphene membranes we did not observe any leakage through the membrane and/or membrane/substrate interface until a critical pressure limit, then a sudden membrane detachment occurs. PG and BPC membranes are not impermeable as graphene ones, but there are significant energy barriers to diffusion depending on the atom type. Our results show that this kind of porous membranes can be effectively used as selective sieves for pure and mixtures of gases.

## INTRODUCTION

Membranes are very important structures for a large variety of scientific and technological applications. They can be used as selective barriers (creating regions with very different physical and chemical properties) and play a very important role in process such as cellular compartmentalization, industrial-scale chemical, gas, water purifications and mechanical pressure sensing [1-4].

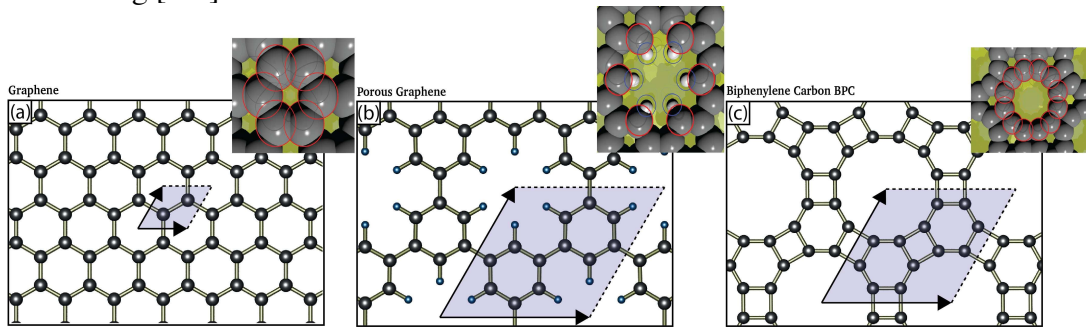

**Figure 1.** Investigated membrane structures. (a) graphene; (b) porous graphene (PG), and; (c) biphenylene carbon (BPC). In the Figure inset the atoms are displayed showing their van der Waals radii values. The shaded areas indicated the structural unit cells.

Graphene (Figure 1-a), carbon one-atom thick membrane, has become one of the hottest topics in materials science today. Besides its unique electronic properties, graphene also exhibits high mechanical strength [5], high thermal conductivity [6] and high optical absorptivity [7]. Recently [1], it was proposed that graphene could act as impermeable atomic membrane to standard gases as He, Ar and $N_2$. However, there are many applications where a higher level of porosity is needed. In part due to this, there is a renewed interest in discovering graphene-like porous membranes. Porous graphene (PG) [8] (Figure 1-b) and biphenylene carbon (BPC) [9,10] (Figure 1-c) are two of these structures. PG has a topology quite similar to graphene, with pores being formed from the replacement of some ring carbon atoms by hydrogen ones. BPC has also a quite interesting topology with its pores resembling typical sieve cavities and/or some kind of zeolites.

In this work we have used molecular dynamics (MD) simulations to study the permeability of the membranes shown in Figure 1 and possible selectivity to some specific compounds for the porous ones. In this work we have considered the cases of He and Ar atoms.

**THEORY**

The MD simulations were carried out mimicking the experimental conditions [1], where a nanochamber composed of amorphous silicon oxide is filled with different compounds and sealed by the membranes. The pressure inside the chamber is then increased and the gas leakage dynamics (through the membranes and between the membranes and the substrates) analyzed. Membranes of dimensions 220 x 220 Å$^2$, 230 x 230 Å$^2$, and 240 x 240 Å$^2$ were used to seal a cavity of 200 x 200 Å$^2$ composed of amorphous silicon oxide slabs. The membranes were kept over the surface only due to van der Waals interactions (up to the limit of detachment) and all atoms set free to move, while the atoms composing the substrates were kept fixed during the simulations. The van der Waals forces are proportional to the contact area. In this work we considered three different contact area (overlap) cases, with arbitrary values of 8400, 17600 and 27600 Å$^2$. In Figure 2 (inset) we present a schematic representation of the membrane covering the cavity. The overlap area is measured from the cavity edge to the membrane edge.

All the calculations were carried out through fully atomistic molecular dynamics simulations using the classical force field (CHARMM [11]), as implemented on LAMMPS software [12]. A thermostat was attached to gas atoms in order to keep the temperature constant at T=300K. There is no thermostat coupled directly to the membranes. Membrane temperatures are controlled indirectly through the direct contact with the gas. In order to increase the gas pressure, a piston placed at bottom of the cavity was used to decrease the gas volume with a constant velocity of $6.0 \times 10^{-2}$ Å/ps. As the piston continuously decreases the gas volume, the gas pressure increases. All the calculations were performed using a time step of 1.0 fs.

For the calculations for the barrier energies to make He and Ar atoms go through the porous (PG and BPC) membranes we considered isolated membranes (no substrates) with their border atoms fixed and the remaining atoms free to move. A probe atom (He or Ar) is placed at 10 Å above one of the membrane porous (aligned with is center). The probe atom is then moved with constant velocity of $1.0 \times 10^{-2}$ Å/ps to pass thorough the porous and the energy configurations calculated. A thermostat is coupled to the membrane atoms allowed to move and temperature is kept low (around T=10K), in order to avoid undesirable large out-of-plane membrane fluctuations.

**DISCUSSION**

In Figure 2 we present the results for the pressure behavior as a function of time for the case of a graphene membrane (200 x 200 Å$^2$) over a nanochamber filled with Ar gas. When the pressure of the Ar gas confined inside the cavity is increased, due to the piston action, the membrane starts to stretch (bulging). For all the cases considered here no gas leakage was observed through the graphene membranes (as expected) or between the graphene membranes and the substrates (consistent with the experimental data [1]), until the pressure value reaches a critical limit and then an abruptly detachment occurs. These pressure limit values can be identified by an abrupt decrease in the pressure values (see Figure 2). This behavior can be explained by a sudden gas expansion (consequently, a pressure value decrease) just after the membrane detaches from the surface. As expected, these pressure limit values are proportional to the overlap area. Similar results (with the exception that gas leakage through the membranes occur) were observed for PG and BPC membranes.

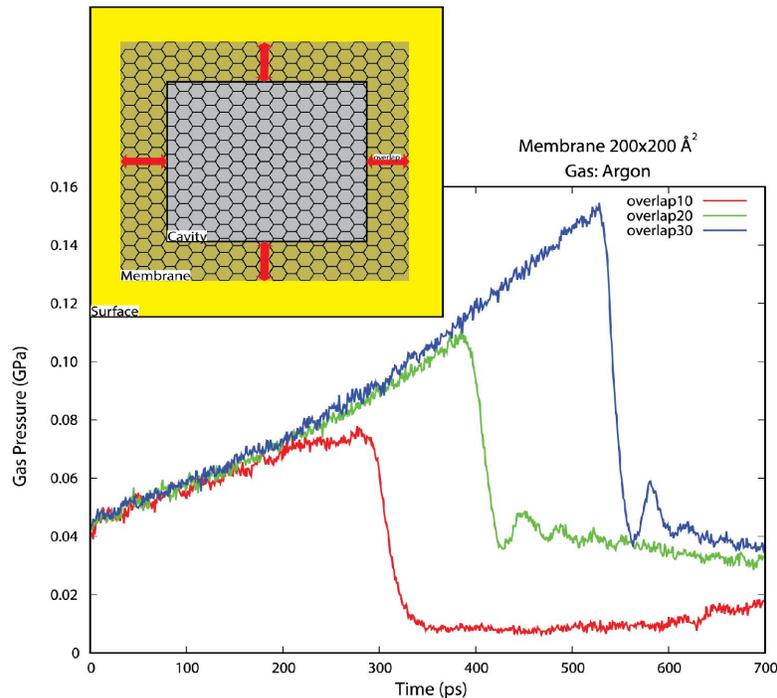

**Figure 2.** Pressure values inside the cavity as a function of time. The discontinuity (abrupt fall) represents the maximum pressure before the membrane detaches from the surface. Overlap10, overlap20 and overlap30 refer to the overlap area values of 8400, 17600 and 27600 Å$^2$, respectively. In the Figure inset is shown a schematic representation of the membrane over the nanochamber pit. The distance between the cavity border and the membrane edge defines the overlap area.

Experimental and theoretical data suggest that it is not possible to pass atoms (even He) through graphene membranes (unless high kinetic energy is used, which damage the structure) [1-4]. Our results are in agreement with these data.

On the other hand the higher porosity (in relation to graphene) of the PG and BPC structures, are suggestive that these membranes are not impermeable to He and Ar atoms, as the graphene ones.

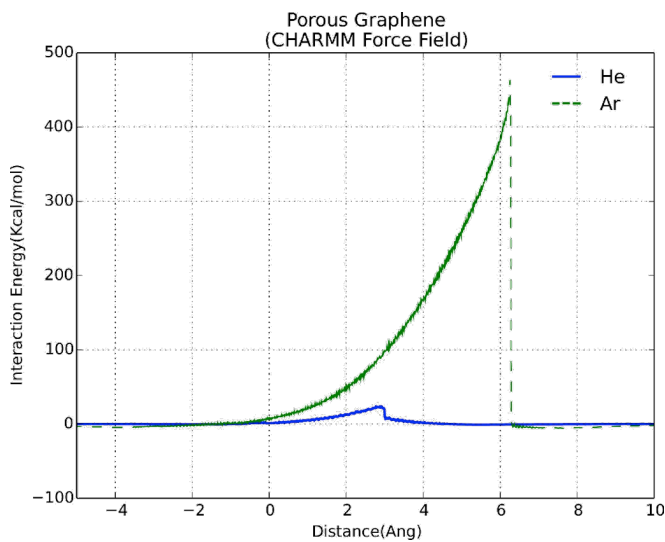

**Figure 3.** Energy barrier to move a probe atom through a porous on the Porous Graphene (PG) membrane. The probe is moved quasi-statically and the interaction energy computed in every step. Membrane edges were kept fixed while the remaining atoms are free to move accordingly to the system forces.

In Figures 3 and 4 we present the dynamic energy barrier to an atom probe (He or Ar) to cross the membranes through one of their pores. The probe atom is moved following a straight path that is perpendicular to the membrane plane and crosses the porous center and the energy configuration calculated at different distances from the membranes.

As the probe atom is moved towards the membrane, they start to interact. The membrane starts to stretch out of plain (see Figure 4) and the porous geometry starts to change. During membrane stretch the porous size increase until there is space enough to the probe atom pass. During this process the interaction energy between the membranes and probes increases. After the atom probe passed through the porous, the membrane starts to restore its original planar configuration. The abrupt drop in energy shown in Figures 3 and 4 can be directly related to this restoration processes. We can associate the maximum energy values as the energies necessary to pass an atom/molecule through the porous.

From Figures 3 and 4 we can see that while it is relatively easy to pass a He atom, the barrier is significantly higher for the Ar case. Interestingly the value for the BPC case is higher than for the PG membrane, although the intrinsic BPC diameter pore is larger than the corresponding PG (5.8 and 3.9 Å, respectively). These results can be explained by the significant differences in the flexibility exhibit by PG and BPC. BPC is a much more rigid structure [9], the presence of the hydrogen atoms in PG makes it more flexible and easily deformed, thus explaining why less energy is needed to cross an Ar atom. The energy profiles reveal different barriers to different atoms, suggestive that this behavior can be exploit to use PG and BPC as selective sieves.

There are many works in the literature regarding calculations for energy barriers of nanomembranes [13]. Some of these calculations were carried out using membranes that are kept frozen during simulations, which in general produce symmetric energy profiles. Allowing the membrane movements during the simulations produce asymmetric energy profiles, as the ones we presented in Figures 3 and 4 and from some *ab initio* calculations [14]. The energy barrier for PG and BPC illustrates the importance of not using frozen membranes in the simulations, what would preclude changes in porous size and form, which could lead to wrong conclusions about the relative easiness of atom diffusion.

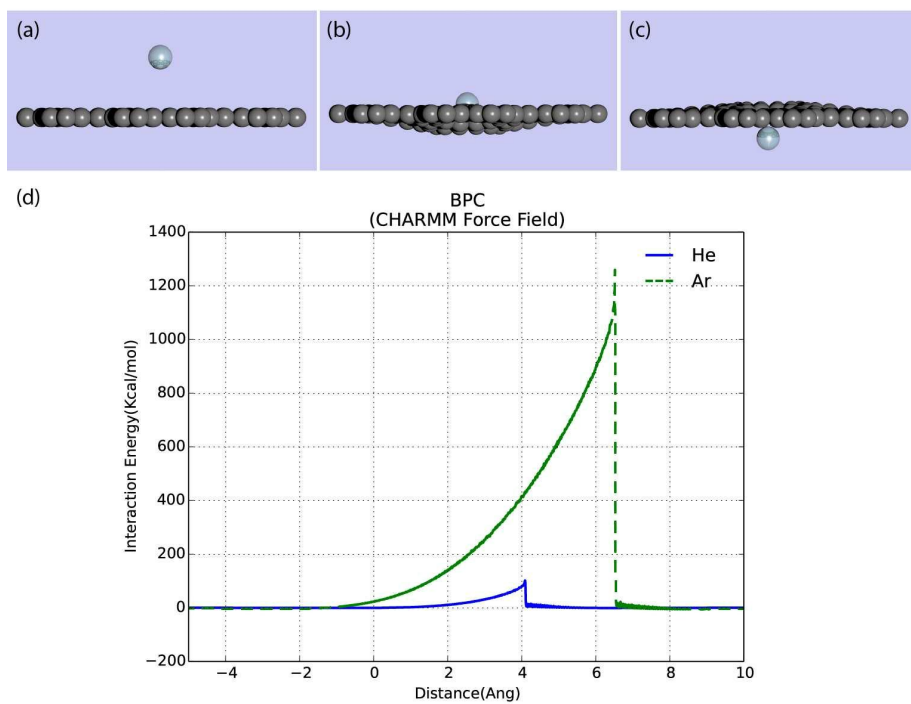

**Figure 4.** (a-c) Snapshots from the atom probe interacting with a biphenylene carbon (BPC) at different distances. The membrane deformations are clearly seen in Figure 4(b). (d) Energy barrier to move a probe atom/molecule through a porous on the BPC membrane. The probe is moved quasi-statically and the interaction energy computed in every step. Membrane edges were kept fixed while the atoms that interact directly with the probe were free to move accordingly to the system forces.

## CONCLUSIONS

We have investigated through fully atomistic molecular dynamic simulations the impermeable (to He and Ar atoms) characteristics of graphene, porous graphene (PG) and biphenylene carbon (BPC) membranes. Our results showed that graphene membranes are impermeable to these gases, even to high pressures (up to 0.15 GPa) . Below this pressure value we did not observe any gas leakage through the membrane structure or through the interface membrane/surface. The absence of leakage between the interface membrane/surface can be explained by considering the high flexibility of graphene [16], which allows conformational rearrangements on corrugated surfaces. We also show that it is possible to tune the maximum supported pressure through controlling the overlap between membrane and substrate. After a critical limit pressure, graphene membranes detaches abruptly form the substrates.

Although PG and BPC exhibited similar behavior to the detachment from the silicon oxide substrates, they are not impermeable membranes to He and Ar gases. Comparatively, PG presents lower resistance to the passage of single atoms in relation to BPC membranes, although the BPC structures present larger intrinsic pores (Figure 1). This behavior can be explained by the intrinsic higher flexibility of PG structures, which allow an easier pore deformation to allow atom diffusion. These characteristics can be exploited to use these membranes in the selective separation of pure gases and/or gases mixtures because their strong structures are capable to stand high-pressure values while allowing the passages of some atoms and blocking the passage of others. Work along these lines are in progress.

**ACKNOWLEDGMENTS**

This work was supported in part by the Brazilian Agencies CNPq, CAPES and FAPESP. The authors thank the Center for Computational Engineering and Sciences at Unicamp for financial support through the FAPESP/CEPID Grant #2013/08293-7. The authors wish also to thank Prof. J. S. Bunch for many helpful discussions.